\documentclass[aps,prb,twocolumn,amsmath,amssymb,superscriptaddress,floatfix]{revtex4}
\usepackage{graphicx} 
\usepackage{bm}
\usepackage[usenames]{color}
\bibstyle{apsrev.bib}

\newcommand{\be}{\begin{equation}}
\newcommand{\ee}{\end{equation}}
\newcommand{\beqn}{\begin{eqnarray}}
\newcommand{\eeqn}{\end{eqnarray}}

\begin{document}

\title{Nonequilibrium dynamics of the Ising chain in a fluctuating transverse field}
\author{Gerg\H o Ro\'osz}
\email{roosz.gergo@wigner.mta.hu}
\affiliation{Wigner Research Centre for Physics, Institute for Solid State Physics and Optics, H-1525 Budapest, P.O.Box 49, Hungary}
\affiliation{Institute of Theoretical Physics, Szeged University, H-6720 Szeged, Hungary}
\author{R\'obert Juh\'asz}
\email{juhasz.robert@wigner.mta.hu}
\affiliation{Wigner Research Centre for Physics, Institute for Solid State Physics and Optics, H-1525 Budapest, P.O.Box 49, Hungary}
\author{Ferenc Igl\'oi}
\email{igloi.ferenc@wigner.mta.hu}
\affiliation{Wigner Research Centre for Physics, Institute for Solid State Physics and Optics, H-1525 Budapest, P.O.Box 49, Hungary}
\affiliation{Institute of Theoretical Physics, Szeged University, H-6720 Szeged, Hungary}
\date{\today}

\begin{abstract}

We study nonequilibrium dynamics of the quantum Ising chain at zero temperature when the transverse field is varied stochastically. In the equivalent fermion representation, the equation of motion of Majorana operators
is derived in the form of a one-dimensional, continuous-time quantum random walk with stochastic, time-dependent transition amplitudes. This type of external noise gives rise to decoherence in the associated quantum walk and the semiclassical wave-packet generally has a diffusive behavior. As a consequence, in the quantum Ising chain, the average entanglement entropy grows in time as $t^{1/2}$ and the logarithmic average magnetization decays in the same form. 
In the case of a dichotomous noise, when the transverse-field is changed in discrete time-steps, $\tau$,
there can be excitation modes, for which coherence is maintained, provided their energy satisfies $\epsilon_k \tau\approx n\pi$ with a positive integer $n$. 
If the dispersion of $\epsilon_k$ is quadratic, the long-time behavior
of the entanglement entropy  and the logarithmic magnetization is dominated by these ballistically traveling coherent modes and both will have a $t^{3/4}$ time-dependence.

\end{abstract}

\pacs{}

\maketitle

\section{Introduction}

Recent progress of experiments with ultracold atoms in optical lattices\cite{Greiner_02,Paredes_04,Kinoshita_04,Kinoshita_06,Lamacraf_06,Sadler_06,Hofferberth_07,bloch,Trotzky_12,Cheneau_12,Gring_11} has triggered intensive theoretical
research to understand the properties of nonequilibrium relaxation process of closed quantum systems. One basic question is related to the behavior of the system after a (global) quantum quench, i.e.
after sudden change of parameters in the Hamiltonian
\cite{Polkovnikov_11,barouch_mccoy,igloi_rieger,sengupta,Rigol_07,Calabrese_06,Calabrese_07,Cazalilla_06,Manmana_07,Cramer_08,Barthel_08,Kollar_08,Sotiriadis_09,Roux_09,Sotiriadis_11,Kollath_07,Banuls_11,Gogolin_11,Rigol_11,Caneva_11,Cazalilla_11,Rigol_12,Santos_11,Grisins_11,Canovi_11,Calabrese_05,Fagotti_08,Silva_08,Rossini_09,Campos_Venuti_10,Igloi_11,Rieger_11,Foini_11,Calabrese_11,Schuricht_12,Calabrese_12,blass,Essler_12,evangelisti_13,fagotti_13,pozsgai_13a,fagotti_essler_13,collura_13,bucciantini_14,fagotti_14,cardy_14,wouters,pozsgay,goldstein,pozsgay1,mestyan,prosen}. After sufficiently long time, the system evolves to a stationary state which is, however different for integrable and nonintegrable systems. Nonintegrable systems generally show thermalization\cite{Rigol_07,Calabrese_06,Calabrese_07,Cazalilla_06,Manmana_07,Cramer_08,Barthel_08,Kollar_08,Sotiriadis_09,Roux_09,Sotiriadis_11}, whereas for integrable systems the so called generalized Gibbs ensemble is expected to hold\cite{wouters,pozsgay,goldstein,pozsgay1,mestyan,prosen}. Concerning the functional form of the relaxation process a few exact results are available for integrable
systems, which can be - even qualitatively - explained in the frame of a semiclassical theory\cite{Calabrese_05,
Calabrese_07,Igloi_11,Rieger_11}. This is based on the observation, that after the quench, entangled pairs of excitations (so called quasi-particles) are emitted, which propagate ballistically (in opposite directions) in translationally invariant 
systems. This explains, among others, the linear increase of the entanglement entropy and the exponential decrease of the magnetization after a global quench in homogeneous chains. This semiclassical theory can be used for nonintegrable systems, furthermore, this picture explains qualitatively the sub-ballistic dynamics in non-homogeneous (random\cite{dyn06,bo,Igloi_12,Levine_12,Pollman_12,Vosk_12} or aperiodic\cite{irl}) systems. For accelerated dynamics, see Refs\cite{eg,ni}.

Besides global quenches, another time-dependent processes have been investigated, as well. Here we mention \textit{local}
quenches\cite{Calabrese_07,stephan_dubail,ep-07,ekpp-08,ISzL09,sodano1,sodano2,Divakaran_11}, when only a few parameters are changed suddenly; \textit{adiabatic relaxation}\cite{kibble,zurek,polkov05,caneva,levitov,dutta,singh,mondal1,mondal2,polkov08,patane,degrandi,bermudez,vish,pollmann,dziarmaga,thakurathi},
when the parameters are slowly (generally linearly) ramped through a quantum critical point, a process also used in quantum annealing for random systems\cite{sandvik}; and {\it periodic quench}\cite{ep,das}, which is a sequence of single quenches occurring at discrete times. 
Periodically driven quantum systems are interesting on their own right\cite{bukov}, and show many unexpected phenomena, like the Kapitza pendulum\cite{kapitza}, which are not present in equilibrium systems.
In this case the Hamiltonian of the system is time dependent and often finite dimensional, which is then
mapped onto an infinite-dimensional but time-independent Floquet Hamiltonian.

In the present paper, we consider a different setup, when the drive is time-dependent, but not periodic, thus Floquet theory does not hold. Namely, the drive of the quantum system under study has a stochastic character; it varies randomly in time but it is perfectly correlated in space, mimicking interaction with a fluctuating environment. 
Similar models, namely one-dimensional nearest-neighbor interacting quantum spin chains with external fields fluctuating independently on each site have recently been studied from the aspect of information propagation\cite{beo}.  
We can then ask the question how an external noise alters the conservative time evolution of observables after a quench.
We aim at studying this question in the Ising chain subject to a globally fluctuating transversal magnetic field. 
Recently, the nonequilibrium dynamics of this model with a weak Gaussian white noise have been studied with a focus on the crossover from a prethermalized regime toward a thermalized one, where the transversal correlator is characterized by a diffusive behavior\cite{ms}.  
The advantage of this model is that the standard free-fermion technique makes possible an efficient numerical treatment of the dynamics.
We will point out a close relationship between the dynamics of this model and continuous-time quantum walks (CTQW)\cite{ctrw,kempe} in the presence of temporal noise. Studies of the latter model have been motivated by understanding of decoherence in quantum systems. Based on numerical results, temporal noise in the CTQW is conjectured to destroy interference manifesting itself in ballistic spreading and to give way for diffusive spreading characteristic for classical random walks for long times\cite{decoherence,kendon}.
Here, we will present numerical results for the time-dependence of the average entanglement entropy and the relaxation of the average magnetization of the Ising chain in a transverse field that is switched randomly between two values at discrete times but remains constant within periods of duration $\tau$. 
Interestingly, the large-frequency (small $\tau$) and the low-frequency (large $\tau$) regimes show different asymptotic behaviors.
In the former case, the associated CTQW, which describes the quasi-particles created after the quench, will lose its quantum coherence due to the temporal noise and spreads diffusively. 
We will argue within a semiclassical theory and confirm by numerical results that this leads to a square-root time-dependence of the entanglement entropy and the logarithmic magnetization. 
For slow enough variations (large $\tau$), however, quantum coherence survives temporal noise for discrete excitations, which therefore still propagate ballistically. 
For certain cases, these rare modes will dominate the dynamics of the above quantities, resulting in a different asymptotic time dependence.

The rest of the paper is organized in the following way. In section \ref{model}, the model is introduced and, using its fermion representation, a relationship to continuous-time random quantum walks is pointed out. 
In section \ref{noneq}, the evolution of different quantities, such as the spatiotemporal correlation function, entanglement entropy and magnetization, are studied numerically and analytically. A theory explaining the deviations from diffusive behavior by the existence of stroboscopic eigenmodes is presented. 
Finally, results are discussed in section \ref{discussion}, and some of the details of calculations are deferred to the Appendix.

\section{Ising dynamics and quantum walks}

\label{model}

We are going to study the spin-1/2 transverse-field Ising chain with time-dependent parameters, defined by the Hamiltonian
\be
{\cal H}(t) =
-\frac{J(t)}{2}\sum_{i=1}^{L-1}\sigma_i^x \sigma_{i+1}^x-\frac{h(t)}{2}\sum_{i=1}^L\sigma_i^z\;,
\label{H}
\ee
where $\sigma_i^x$ and $\sigma_i^z$ are Pauli operators at site $i$. The number of sites $L$ is assumed to be even. 
Note that, for the sake of concreteness and simplicity in numerical calculations, we have chosen here free boundaries, but our asymptotic results apply to the bulk of a large system, where boundary effects do not play a role.  

We consider a simple form of time-dependence with piecewise constant Hamiltonians in periods of duration $\tau$. The Hamiltonian ${\cal H}(n)$ acting in the $n$th time interval $(t_{n-1},t_n]$, where $t_n\equiv n\tau$, $n=1,2,\dots$, is chosen randomly from a set 
$\{{\cal H}_l\}_{l=1}^N$ of non-commuting, constant Hamiltonians ${\cal H}_l$ containing parameters $J_l$ and $h_l$. In the numerical calculations we used a dichotomous noise ($N=2$), 
where one of two Hamiltonians is chosen at the beginning of each period independently with equal probabilities. The parameters we mainly used were $J_1=J_2=1$, $h_1=h$, $h_2=-h$\footnote{Note that, although with this choice both Hamiltonians are critical, this is not a substantial property as the phenomena to be described in the sequel 
emerge irrespective of whether the Hamiltonians are critical or not.}.  

We considered then the unitary time evolution from some initial state $|\Psi_0\rangle$: 
\be 
|\Psi(t_n)\rangle={\cal U}_n{\cal U}_{n-1}\cdots {\cal U}_1|\Psi_0\rangle,
\ee
where ${\cal U}_n=e^{-{\rm i}{\cal H}(n)\tau}$. 
Owing to the simple choice of a piecewise constant time-dependence of the Hamiltonian, the time evolution is composed of a sequence of conservatively evolving segments. Let us therefore first recapitulate the nonequilibrium dynamics with a constant Hamiltonian, i.e. $J(t)=J$, $h(t)=h$, and then write it in a form most comfortable for constructing time evolution in the noisy model. 

As it is well known, the Hamiltonian in Eq. (\ref{H}) can be written in a quadratic form of fermion creation
($c_i^{\dagger}$) and annihilation ($c_i$) operators by means of the Jordan-Wigner transformation\cite{lieb61} as 
\be
{\cal H}=
-\frac{J}{2}\sum_{i=1}^{L-1}(c_i^{\dagger}-c_i)(c_{i+1}^{\dagger}-c_{i+1})
 -h\sum_{i=1}^L(c_i^{\dagger}c_i-\frac{1}{2})\;.
\label{Hfermion}
\ee
In terms of Clifford operators defined as 
\beqn
\hat{d}_{2i-1}=c_i^{\dagger}+c_i=(\prod_{j<i}-\sigma_j^z)\sigma_i^x, \nonumber \\ 
\hat{d}_{2i}=c_i^{\dagger}-c_i=i(\prod_{j<i}-\sigma_j^z)\sigma_i^y, \nonumber \\
i=1,2,\dots,L
\eeqn
and having the anticommutation relations 
\be 
\{\hat{d}_m,\hat{d}_n\}=2(-1)^{m-1}\delta_{mn}
\label{anti}
\ee
the Hamiltonian assumes the form
\be
{\cal H} =\frac{1}{4}\sum_{i,j=1}^{2L}\hat{d}_i^{\dagger}H_{ij}\hat{d}_j
\label{HClifford}
\ee
with the symmetric matrix
\be 
H=
\begin{pmatrix}
0   & h&               &        &               &    \cr
h   &0            & J &        &               &    \cr
    & J          &  0           & h &         &    \cr  
    &               &  h         & 0     & \ddots        &    \cr
    &               &               & \ddots & \ddots       & h \cr  
    &               &               &        &  h          & 0 
\end{pmatrix}.
\label{Tmatrix}
\ee
Using the relations in Eq. (\ref{anti}), one obtains the equation of motion of Clifford operators in Heisenberg picture in the form 
\be 
\frac{d\hat{d}_i(t)}{dt}=-{\rm i}\sum_{j=1}^{2L}H_{ij}\hat{d}_j(t).
\label{eom}
\ee
This form of the evolution equations makes the relationship of the model with continuous-time quantum random walks transparent. Clearly, an identical form of equations can be written for the matrix element of $\hat{d}_i(t)$ between two fixed states as for $\hat{d}_i(t)$ itself in Eq. (\ref{eom}). 
These equations can then be interpreted as a CTQW on a one-dimensional open lattice with $2L$ sites and with alternating transition amplitudes $h$ and $J$ on odd and even bonds, respectively, while the properly normalized matrix elements of $\hat{d}_i(t)$ play the role of probability amplitudes of the quantum walk at time $t$.

Before proceeding with the stochastic model, two caveats are in order. First, the relation between the dynamics of the quantum Ising chain and the CTQW holds also for inhomogeneous systems, in which the couplings, $J_i$, and the transverse fields, $h_i$, are position dependent. Being the dynamics of the random transverse-field Ising chain
ultra-slow, the same should be true for the CTQW with spatial disorder. Second,  the symmetric matrix in Eq. (\ref{Tmatrix}) can be interpreted as the transfer matrix of a classical, discrete-time random walk model. This correspondence has been used to connect the equilibrium critical behavior of the quantum Ising chain and that of the related classical random walk\cite{igloi_turban,igloi_rieger_bigp}.

Let us now return to the time-dependent model with noise. We will restrict ourselves to a stroboscopic view of the time evolution at discrete times $t_n=n\tau$, $n=0,1,2,\cdots$. This facilitates numerical calculations since one only needs to calculate the unitary evolution matrices 
$U_1=e^{-{\rm i}H_1\tau}$ and $U_2=e^{-{\rm i}H_2\tau}$ over periods $\tau$ with constant Hamilton matrices $H_1$ and $H_2$, respectively. This can easily be done via diagonalising $H_1$ and $H_2$, which have the form given in Eq. (\ref{Tmatrix}). The resulting matrices $U_1$ and $U_2$ contain complex entries, see e.g. Ref.\cite{igloi_rieger}, but working with self-adjoint Majorana operators
\be 
\check{a}_{2i-1}=\hat{d}_{2i-1}, \qquad 
\check{a}_{2i}=-{\rm i}\hat{d}_{2i}, \quad i=1,2,\dots,L
\ee
rather than with $\hat{d}_i$, their evolution matrices over $\tau$, $O_1$ and $O_2$ will be real, see Ref. \cite{ISzL09}. 
In the noisy system, as we defined above, the evolution matrix $O(n)$ in the $n$th period is either $O_1$ or $O_2$ with equal probabilities. Note, that between $\check{a}_i(\tau)$ and its initial values, $\check{a}_j$, the matrix, $O(1)$, represents a linear relation. After $n$ steps
for a given realization of the temporal noise, the time evolution is thus given by 
\be 
\check{a}_i(t_n)=\sum_{j=1}^{2L}[O(1)O(2)\cdots O(n)]_{ij}\check{a}_j.
\label{unitary}
\ee
Working again with Clifford operators, their time evolution in the noisy system is equivalent with a noisy CTQW, in which the transition amplitudes change randomly at discrete times $t=t_n$. 

\section{Nonequilibrium relaxation in temporal noise}

\label{noneq}
\subsection{Spatiotemporal correlation function}
In the framework of a semiclassical theory mentioned in the Introduction, 
the key to the understanding of the nonequilibrium dynamics in case of sudden quenches is the way an excitation propagates through the chain. This can be characterized by the correlation function of Majorana operators
\be 
G_l(t)=\frac{1}{\sqrt{2}}\langle x|\check{a}_l(t)\check{a}_L|x\rangle,
\label{Gnt}
\ee
where $|x\rangle$ denotes the product state polarized in the positive $x$ direction, $|x\rangle\equiv|\rightarrow\rightarrow\dots\rightarrow\rangle$. 
\footnote{Note that, in Ref.\cite{irl}, the correlation function $\langle c_l(t)c_{L/2}^{\dagger}\rangle$ of fermion operators (formulated as a wave packet) has been studied, rather than $G_l(t)$. While this has a long-time behavior similar to that of $G_l(t)$, the latter has the advantage of remaining precisely normalized in the sense of Eq. (\ref{norm}). A similar conservation of the norm of $\langle c_l(t)c_{L/2}^{\dagger}\rangle$ is valid in the hopping (XX) model.} 
It is easy to calculate that the only non-zero initial values of $G_l(t)$ at time $t=0$ are $G_L(0)=1/\sqrt{2}$ and $G_{L+1}(0)={\rm i}/\sqrt{2}$ and, due to the unitary evolution for a given realization of the noise [see Eq. (\ref{unitary})], 
\be
\sum_{l=1}^{2L}|G_l(t)|^2=const=1
\label{norm}
\ee
for any $t\ge 0$.

The correlation function in Eq. (\ref{Gnt}) has a clear interpretation in the quantum walk picture. 
Up to a constant phase factor, it corresponds to the amplitude $A_l(t)$ on site $l$ at time $t$ of a quantum walk, which was initialized in the middle of the chain, i.e. on sites $L$ and $L+1$ with amplitudes $-1/\sqrt{2}$: 
\be 
A_l(t)\equiv \frac{1}{\sqrt{2}}\langle x|\hat{d}_l(t)\hat{d}_L|x\rangle=
-(-{\rm i})^pG_l(t), 
\ee
where $p=l\mod 2$. With this initial condition, the probability distribution of the position of the walker will be left-right symmetric, i.e. $|G_l(t)|^2=|G_{L-l+1}(t)|^2$.

In order to probe the dynamics of quasiparticle excitations, we have 
numerically studied the time evolution of the probabilities $|G_l(t)|^2$ in the model with temporal disorder. 
Clearly, $|G_l(t)|^2$, similarly to other observables, will depend on the particular realization of the noise, and in case of a sufficiently narrow distribution, which is the case here according to numerical results, it is reasonable to consider an average over different temporal histories. 
A considerable advantage of the chosen dichotomous noise is that the average of certain observables such as $|G_l(t)|^2$ can be efficiently computed by writing appropriate recursions directly for the average. 
To see this, let us start with the time evolution of the matrix
\be 
G_{ij}(t)\equiv G_i^*(t)G_j(t)
\ee 
during a single segment: 
\be 
G_{ij}(t_n)=\sum_{k,l=1}^{2L}O_{ik}(n)O_{jl}(n)G_{kl}(t_{n-1}),
\ee
where we have used that $O(n)$ is real. 

Performing now the averaging results in 
\beqn 
\overline{G_{ij}(t_n)}=\sum_{k,l=1}^{2L}\overline{O_{ik}(n)O_{jl}(n)}\cdot\overline{G_{kl}(t_{n-1})}= \nonumber \\
\sum_{k,l=1}^{2L}\frac{1}{2}\{[O_1]_{ik}[O_1]_{jl}+[O_2]_{ik}[O_2]_{jl}\}\overline{G_{kl}(t_{n-1})}.
\label{directav}
\eeqn
Here, we have made use of the fact that the temporal noise is uncorrelated. 
Using the second line in Eq. (\ref{directav}), $\overline{G_{ij}(t_n)}$
can be computed recursively starting from the initial values  $\overline{G_{ij}(0)}$ by $O(L^3n)$ operations.  The diagonal elements $\overline{G_{ii}(t_n)}$ are the probabilities we are looking for. 
We have numerically calculated $\overline{|G_{l}(t_n)|^2}$ and, the corresponding CTQW being dimerized, we have considered the probability $p_l(t)$ of the walker being in the $l$th cell comprising site $2l-1$ and $2l$: 
\be 
p_l(t)\equiv\overline{|G_{2l-1}(t)|^2}+\overline{|G_{2l}(t)|^2}, 
\qquad l=1,2,\dots,L.
\ee
as well as the variance: 
\be 
\sigma^2(t)=\sum_{l=1}^L\left(l-l_0\right)^2p_l(t),
\label{var}
\ee
where $l_0=\frac{L+1}{2}$.

\begin{figure}
\includegraphics[width=8cm]{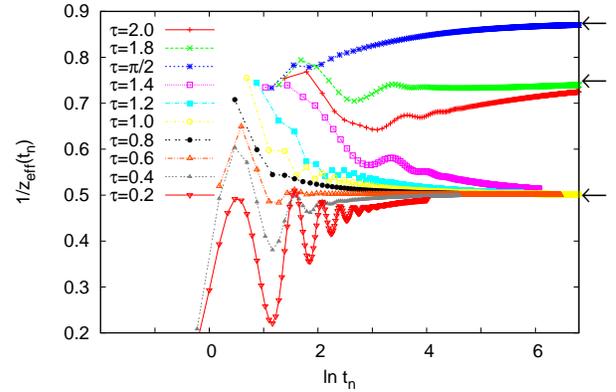}
\caption{\label{fig_var} (Color online) Numerically calculated inverse dynamical exponent as defined in Eq. (\ref{invexp}) as a function of time for $J_1=J_2=1$, $h_1=-h_2=1$ and different $\tau$. The topmost data correspond to $\tau=\pi/2$, while the other data from bottom to top correspond to values of $\tau$ in an increasing order. 
The size of the system is $L=2048$. The arrows indicate the asymptotic values obtained by the theory in section \ref{theor_strob}.}
\end{figure}

We have performed numerical calculations keeping the couplings time-independent, $J_1=J_2=1$, and letting only the sign of the transverse field fluctuate, i.e. $h_1=-h_2=h$ for different $h$ and $\tau$. Expecting a power-law dependence of the variance on time, 
\be 
\sigma^2(t)\sim t^{2/z},
\ee
for long times, we have calculated an effective, inverse dynamical exponent 
\be 
\frac{1}{z_{\rm eff}(t_n)}=\frac{\ln[\sigma(t_n)/\sigma(t_{n-1})]}{\ln(t_n/t_{n-1})}
\label{invexp}
\ee
from finite-time data. These are plotted against time in Fig. \ref{fig_var} for $h=1$ and various $\tau$. At the times used in the numerical calculations, the finite-size effects coming from the finiteness of the lattice are negligible.


As can be seen in the figure, the increase of the variance is slower-than-ballistic, i.e. $1/z_{\rm eff}(t)$ seems to tend to a limiting value $1/z(\tau)$, which is less than one and is dependent on $\tau$.  
For fast enough variations of the magnetic field, $\tau<\pi/2$, the effective inverse dynamical exponent seems to approach $1/2$, although the convergence is slow for  $\tau\lesssim\pi/2$. This tendency changes abruptly at $\tau=\pi/2$. From this value on, $\tau\ge \pi/2$, $1/z(t)$ seems to tend to values definitely higher than $1/2$. The highest value is observed right at the edge of this domain, $\tau=\pi/2$, as well as for $\tau=\pi$ (not shown), while, for $\tau>\pi/2$, the limiting values are somewhat lower. 

\begin{figure}
\includegraphics[width=8cm]{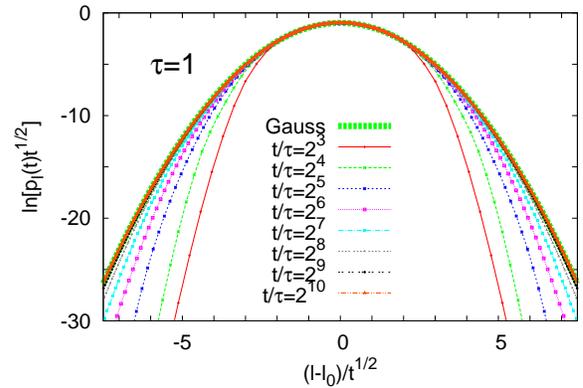}
\caption{\label{fig_p1} (Color online) 
Numerically calculated distribution $p_l(t)$ rescaled according to Eq. (\ref{diff}) at different times. The size of the system is $L=512$, the parameters of the model are $J_1=J_2=1$, $h_1=-h_2=1$ and $\tau=1$. The thick curve is a Gaussian fitted to the data at $t/\tau=2^{10}$}
\end{figure}

The distributions $p_l(t)$ also look differently for short and long $\tau$. 
For short enough $\tau$, as exemplified in Fig. \ref{fig_p1} for $\tau=1$, 
the profile shows diffusive scaling 
\be 
p_l(t)=t^{-1/2}\tilde p[(l-l_0)t^{-1/2}],
\label{diff}
\ee 
where the scaling function $\tilde p(x)$ fits well to a Gaussian. 

\begin{figure}
\includegraphics[width=8cm]{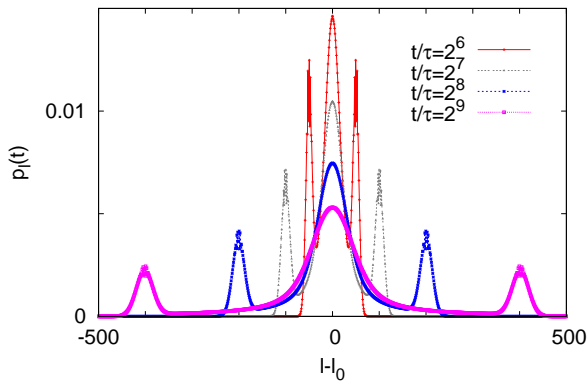}
\caption{\label{fig_pph+} (Color online) Numerically calculated probability distribution $p_l(t)$ at different times. The size of the system is $L=2048$, the parameters of the model are $J_1=J_2=1$, $h_1=-h_2=1$ and $\tau=\pi/2$.}
\end{figure}

\begin{figure}
\includegraphics[width=8cm]{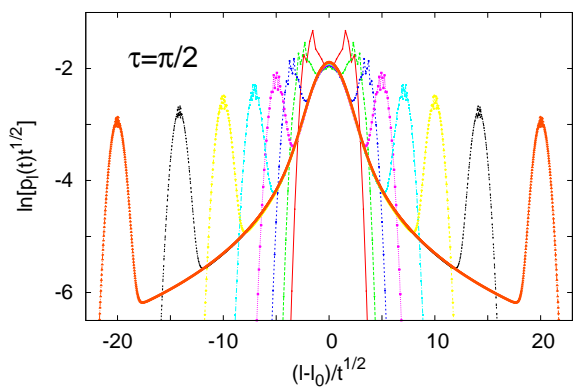}
\caption{\label{fig_pph} (Color online) Scaling plot of the probability distributions $p_l(t)$ at different times shown in Fig. \ref{fig_pph+}.}
\end{figure}

\begin{figure}
\includegraphics[width=8cm]{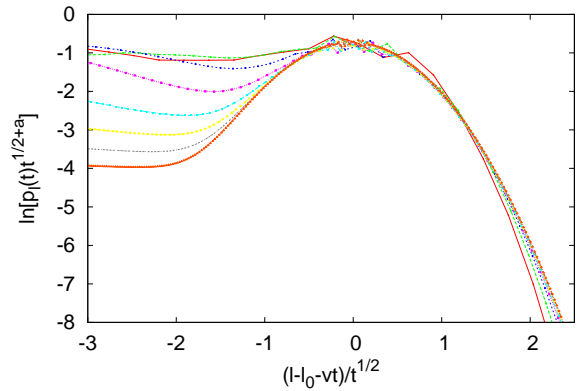}
\caption{\label{fig_pph++} (Color online) Scaling plot of the probability distributions around the right-moving peak for the same parameters as in Fig. \ref{fig_pph+}. The velocity of the center of the peak is $v=0.5$ and the scaling exponent of the weight decrease is $a=0.3$.}
\end{figure}

For larger $\tau$, $\tau\sim 1.4$, the diffusive scaling still holds but the scaling function starts to deviate from a Gaussian, possessing a slower decaying tail. 
At $\tau=\pi/2$, as shown in Figs. \ref{fig_pph+} and \ref{fig_pph}, 
the central part of the profile shows diffusive scaling, but two symmetrically placed peaks appear, which move outwards ballistically. They spread out diffusively and, at the same time, continuously lose their weight $W(t)$ as 
\be 
W(t)\sim t^{-a},
\ee
with $a=0.30$ at $\tau=\pi/2$, see the scaling plot of the part of the profile around a peak in Fig. \ref{fig_pph++}.  


\begin{figure}
\includegraphics[width=8cm]{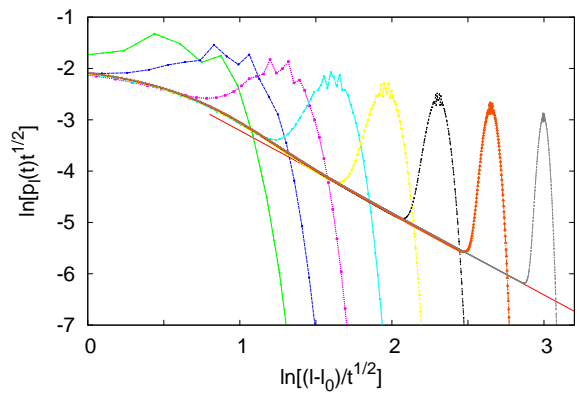}
\caption{\label{fig_pph+++} (Color online) Log-log plot of the scaled probability distributions for the same parameters as in Fig. \ref{fig_pph+}. The slope of the straight line is $-1.60$.}
\end{figure}

It is easy to see that the slow decrease of the weight of the peaks leads to the buildup of fat (algebraic) tails of the scaling function $\tilde p(x)$. Let us consider the total probability outside of a ballistically expanding domain:
\be 
P_>(t)=\sum_{|l'|>vt}p_{l'}(t), 
\ee
where $l'=l-l_0$ and assume that $v$ is smaller than the velocity of the peak. 
We have then $P_>(t)\sim t^{-a}$ or, equivalently, in terms of the scaling variable $x=l'/\sqrt{t}=v\sqrt{t}$, $P_>(x)\sim (x/v)^{-2a}$. The scaling function $\tilde p(x)=-dP_>(x)/dx$ thus decays as 
\be 
\tilde p(x)\sim x^{-(1+2a)},
\ee
see Fig. \ref{fig_pph+++}.

\begin{figure}
\includegraphics[width=8cm]{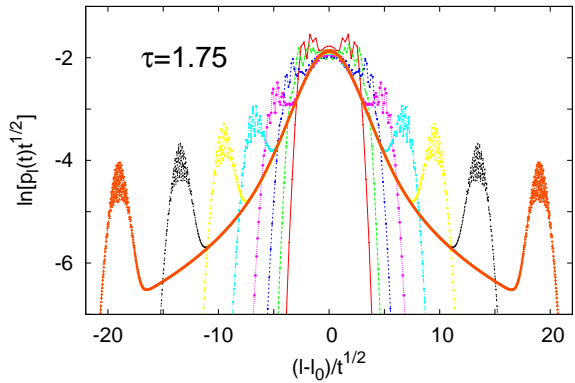}
\caption{\label{fig_1.75} (Color online) Scaling plot of the numerically calculated probability distribution $p_l(t_n)$ at different times. The size of the system is $L=2048$, the parameters of the model are $J_1=J_2=1$, $h_1=-h_2=1$ and $\tau=1.75$.}
\end{figure}

We have computed the probability distributions for other values of the duration $\tau=1.75,2,\pi/\sqrt{2}$ and obtained qualitatively similar results, however, with a velocity of the front dependent on $\tau$ and a decay exponent $a=0.55$ for all the above values, see Fig. \ref{fig_1.75}. 


The contribution of ballistically propagating peaks to the variance $\sigma^2(t)$ is $O(t^{2-a})$, and this suppresses the diffusive contribution of order $t$, since $a<1$. Thus, for $\tau \ge \pi/2$, we obtain $z=2/(2-a)$; otherwise the variance grows linearly with time, i.e. $z=2$. 

\subsection{Theory of stroboscopic eigenmodes}
\label{theor_strob}

In the following, we give an explanation of the anomalous behavior, i.e. different from classical diffusion observed in the case $\tau \ge \pi/2$. 
Let us return to a formulation of the model that is a bit more general than the example of the previous section and consider an arbitrary value of the transverse field, $h$, (the coupling is set to $J=1$ throughout). 
Since boundary effects do not play a role in the spreading of the quasiparticle excitations seen in the numerics, we can concentrate on the bulk of a large system and can make use of translational invariance in finding the eigenmodes of the corresponding CTQW, i.e. eigenvectors of the matrix in Eq. (\ref{Tmatrix}) (for a fixed $h$) in the bulk. Due to the dimerization, these are of two kinds, $f_k^{+}$ and $f_k^{-}$, having the components 
\beqn
f_k^{\pm}(2n-1)&=&\mathcal{N}e^{{\rm i}\Theta_k}e^{{\rm i}kn}, \nonumber \\ 
f_k^{\pm}(2n)&=&\pm \mathcal{N}e^{-{\rm i}\Theta_k}e^{{\rm i}kn},
\eeqn
where $\mathcal{N}=\frac{1}{2}L^{-1/2}$, $\tan 2\Theta_k=-\frac{\sin k}{h+\cos k}$, and the possible wave numbers $k$ ($L$ in number) fill the Brillouin zone $[-\pi,\pi]$ equidistantly.
The corresponding eigenvalues are 
\be 
\epsilon_k^{\pm}=\pm\sqrt{1+h^2+2h\cos k}.
\label{epsilon}
\ee 
Note that, these are nothing but the excitation energies of free fermions $\hat{\eta}_{2k-1}, \hat{\eta}_{2k}$, $k=1,2, \dots,L$  obtained by a Bogoliubov transformation, in terms of which 
the Hamiltonian in Eq.(\ref{HClifford}) assumes a diagonal form:
\be
{\cal H} =\frac{1}{4}\sum_{k=1}^{L}[\epsilon^-_k\hat{\eta}_{2k-1}\hat{\eta}_{2k}+
\epsilon^+_k\hat{\eta}_{2k}\hat{\eta}_{2k-1}]\;.
\label{HClifford_free}
\ee
These modes can then be regarded as the excitations, or quasi-particles that propagate in the system after a quench.

Returning to the corresponding CTQW, let us consider two Hamiltonians $H_A$ and $H_B$, with different transverse fields $h_A$ and $h_B$, respectively, and denote their eigenvectors by $f_{k,A}^{\pm}$ and $f_{k,B}^{\pm}$, respectively. 
The two sets of eigenvectors are related to each other simply as
\be 
f_{k,A}^{\pm}=\frac{1}{2}(e^{{\rm i}\Delta_k}\pm e^{-{\rm i}\Delta_k})f_{k,B}^{+} + 
\frac{1}{2}(e^{{\rm i}\Delta_k}\mp e^{-{\rm i}\Delta_k})f_{k,B}^{-},
\label{relAB}
\ee
where $\Delta_k=\Theta_k^A-\Theta_k^B$.
Consider now the time evolution of eigenmodes of $H_A$ under the action of $U_B(\tau)=e^{-{\rm i}H_B\tau}$: 
\beqn 
U_B(\tau)f_{k,A}^{\pm}=e^{-{\rm i}\epsilon^{+}_{k,B}\tau}\frac{1}{2}(e^{{\rm i}\Delta_k}\pm e^{-{\rm i}\Delta_k})f_{k,B}^{+} + \nonumber \\ 
+e^{{\rm i}\epsilon^{+}_{k,B}\tau}\frac{1}{2}(e^{{\rm i}\Delta_k}\mp e^{-{\rm i}\Delta_k})f_{k,B}^{-},
\label{UBf}
\eeqn
where we have used that $\epsilon_k^{-}=-\epsilon_k^{+}$. 
It is obvious from Eq. (\ref{relAB}), that $f_{k,A}^{\pm}$ are not eigenvectors of $H_B$ since the eigenvalues corresponding to $f_{k,B}^{+}$ and $f_{k,B}^{-}$ are different (having opposite signs).  
Nevertheless, if, for some $k$, the condition 
\be 
\epsilon^{+}_{k,B}\tau=m_B\pi
\label{conditionA}
\ee
is fulfilled with some integer $m_B=1,2,\cdots$, then the pair of vectors $f_{k,A}^{\pm}$ will be eigenvectors of 
$U_B(\tau)$ with eigenvalues $+1$ or $-1$, 
\be
U_B(\tau)f_{k,A}^{\pm}=(-1)^{m_B}f_{k,A}^{\pm}, 
\ee
although they are not eigenvectors of $H_B$ itself. 

The pair of vectors $f_{k,A}^{\pm}$ will thus be common eigenvectors of 
$U_A(\tau)$ and $U_B(\tau)$. 
Similarly, another set of common eigenmodes appear if there exists a wave number $k$ for which
\be 
\epsilon^{+}_{k,A}\tau=m_A\pi
\label{conditionB}
\ee
is fulfilled with some integer $m_A=1,2,\cdots$. 

Restricting the state space to the subspace of such ``stroboscopic'' eigenmodes  (SE) of $H_A$ and $H_B$, the evolution matrices $U_A(\tau)$ and $U_B(\tau)$ will commute, although they are non-commuting on the complete state space. The existence of stroboscopic eigenmodes will appear in the spreading of the quantum walker of as a pair of ballistically moving peaks. 

Note that, if $H_A$ is critical ($h_A=1$), the excitation energy corresponding to the wave number $k=\pi$ will be zero, $\epsilon_{\pi,A}^+=0$. 
Consequently, the eigenmode of $H_B$ corresponding to $k=\pi$
does not change under the action of $U_A(\tau)$, therefore it will be a trivial common eigenmode of $U_A(\tau)$ and $U_B(\tau)$. 
Formally, this corresponds to that Eq. (\ref{conditionB}) is fulfilled with $m_A=0$ for an arbitrary $\tau$. 
But, according to our assumption $[H_A,H_B]\neq 0$, consequently $h_B\neq 1$, and, in this case, one can see from Eq. (\ref{epsilon}), that the group velocity at $k=\pi$ is zero, i.e. $\frac{d\epsilon_{k,B}^+}{dk}|_{k=\pi}=0$, when $U_B(\tau)$ acts. 
Therefore, the velocity of the signal appearing in the wave function due to
the existense of such a trivial SE is zero, i.e. it stays at the origin and thus basically differs from other non-trivial SE-s that propagate with a finite velocity. 
In the case when also the other Hamiltonian is critical ($h_B=-1$), another trivial SE appears at the wave number $k=0$. 

As the conditions in Eqs. (\ref{conditionA}) or (\ref{conditionB}) are met only for a discrete set of wave numbers $k$, a packet built from modes of wave numbers in a narrow range $[k-\Delta k,k+\Delta k]$ around a stroboscopic eigenmode will be gradually losing its weight as components with even a slightly different wave number will not be common eigenmodes and are being scattered out. 

The numerical results obtained for the model $h_A=1$, $h_B=-1$ are in accordance with the above picture. The highest eigenvalue being $\epsilon^{+}_{0,A}=\epsilon^{+}_{\pi,B}=2$, there exist no (non-trivial) stroboscopic eigenmodes for $\tau<\pi/2$. The conditions in Eqs. (\ref{conditionA}) and (\ref{conditionB}) are first met at $\tau=\pi/2$ with $m_A=m_B=1$ for the highest excitation energy, and for $\tau>\pi/2$ with a lower excitation energy $\epsilon=\pi/\tau$. For $\tau\ge m\pi/2$, with integers $m>1$, further stroboscopic eigenmodes appear with excitation energies $\epsilon=m\pi/\tau$. In this region, multiple peaks are expected to emerge in the profile; numerically calculated distributions for $\tau=\pi$ (not shown) indeed contain two distinct peaks moving with different velocities.

\begin{figure}
\includegraphics[width=8cm,angle=0]{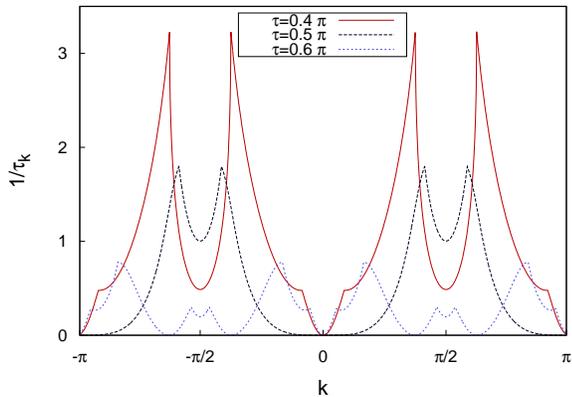}
\caption{\label{relax} (Color online) Inverse lifetime of modes as a function of the wave number $k$ for the model with $J_1=J_2=1$, $h_1=-h_2=1$, for different $\tau$. Note that $1/\tau_k=0$ for $k=0,\pi$ and for arbitrary $\tau$, which is the consequence of the appearance of trivial (non-propagating) SE-s since both Hamiltonians are critical, see the text. For $\tau=\pi/2$, non-trivial (propagating) SE-s with a quadratic dispersion appear at $k=0,\pi$, while for $\tau>\pi/2$, two pairs of non-trivial SE-s with a linear dispersion appear at some intermediate wave numbers.}
\end{figure}

After having explained the origin of ballistically propagating peaks, let us give a quantitative characterization of their rate of decay. In order to do this, let us assume that the initial state is an eigenstate of 
$H_A$, $|f^+_{k,A}\rangle$, and consider the average probability of the system being in this state at time $t$, 
\be 
P_k^+(t)\equiv \overline{|\langle f^+_{k,A}|f(t)\rangle|^2}.
\label{Pk+}
\ee
As it is shown in Appendix \ref{app:decay}, this state decays exponentially as 
\be 
P_k^+(t_n)-P_k^+(\infty)\sim e^{-t_n/\tau_k},
\ee
where the lifetime $\tau_k$  can be obtained as a root of a cubic equation.
The inverse lifetime is plotted against $k$ in Fig. \ref{relax}.  


If, for a fixed $\tau$ and for some $k$, $|f^+_{k,A}\rangle$ (and $|f^-_{k,A}\rangle$) are stroboscopic eigenmodes of $H_B$, then $U_B(\tau)$ in Eq. (\ref{Umatrices}) will be the unit matrix (up to a possible sign), and we have $1/\tau_k=0$. The leading-order dependence of $W(t)$ on time comes from the contribution of slowly relaxing modes around the minima $\tau_{k_0}^{-1}=0$. 

Considering an initial state that is localized in space, it will be a combination of all eigenmodes of $H_A$ with weights that can be taken as constant in the vicinity of $k_0$. We can then write for the contribution of a minimum at $k=k_0$
\beqn
W_{k_0}(t)\sim \int_{k_0-\Delta k}^{k_0+\Delta k}e^{-t/\tau_k}dk \sim  \nonumber \\
\sim \int_{k_0-\Delta k}^{k_0+\Delta k} e^{-tC(k-k_0)^{n_{k_0}}}dk\sim t^{-1/n_{k_0}},
\eeqn
where we have inserted the leading term of the expansion of the inverse lifetime around $k_0$, $\tau_k^{-1}=C(k-k_0)^{n_{k_0}}+o[(k-k_0)^{n_{k_0}}]$.

The order $n_{k_0}$ of the first correction is dependent on whether or not $\epsilon_{k_0,B}^+$ is at the band edge. The reason for this is that, regarding $\tau_{k}^{-1}$ as a function of $\epsilon_{k,B}^+$ rather than of $k$, it is quadratic around $k=k_0$, $\tau_{k}^{-1}\sim (\epsilon_{k,B}^+-\epsilon_{k_0,B}^+)^2$ at any minimum $k_0$. But the dispersion is linear within the band, $\epsilon_{k,B}^+\sim k-k_0$, while it is quadratic at the band edges, $\epsilon_{k,B}^+\sim (k-k_0)^2$, thus we have $n_{k_0}=2$ and $n_{k_0}=4$ in the two cases, respectively. 

If there exists a stroboscopic eigenmode at the band edge (highest excitation energy), we have  therefore $W(t)\sim t^{-1/4}$, while if there are only SE-s with energies within the band, we have $W(t)\sim t^{-1/2}$.
For the model studied numerically, the upper band edge is at $\epsilon^{+}_{0,A}=\epsilon^{+}_{\pi,B}=2$, and such special SE-s exist for the values $\tau=m\pi/2$, $m=1,2,\dots$. In this case we observed $a=0.30$, while otherwise we have seen $a=0.55$. These are slightly higher than the theoretical values $1/4$ and $1/2$, respectively, and the discrepancy may be attributed to corrections to the asymptotic behavior still present at the time scale of numerical calculations.

\subsection{Entanglement entropy}

The nonequilibrium dynamics of the entanglement entropy of a subsystem, which is part of an isolated quantum system contains information about the properties of the excitations which are created after a quench. Here, we divide our system into two halves, thus the subsystem contains the set of spins $n\le L/2$. For simplicity, at $t=0$, the complete system is in a product state $|\psi_0\rangle=|z\rangle\equiv |\uparrow\uparrow\dots\uparrow\rangle$, where all spins point to the positive $z$ direction, thus initially there is no entanglement between the two parts of the system. The entanglement is quantified by the von Neumann entropy
\be 
S(t)=-{\rm Tr}_{n\le L/2}\rho_S(t)\ln\rho_S(t)
\ee
of the reduced density operator $\rho_S(t)={\rm Tr}_{n>L/2}|\psi(t)\rangle\langle\psi(t)|$ of the subsystem at time $t$. For free-fermion models, the entanglement entropy can be calculated from the reduced correlation matrix of Majorana operators, $C_{mn}(t)=\langle \psi_0|\check{a}_m(t)\check{a}_n(t)|\psi_0\rangle$, $m,n=1,\dots,L$ \cite{vidal,ISzL09}. Writing it as $C_{mn}(t)=\delta_{mn}+\Gamma_{mn}(t)$, $S(t)$ is determined by the eigenvalues $\pm \nu_n$, $n=1,\dots,L/2$ of the matrix $\Gamma$ as 
\be 
S(t)=-\sum_{n=1}^{L/2}\left[\frac{1+\nu_n}{2}\ln\frac{1+\nu_n}{2}+\frac{1-\nu_n}{2}\ln\frac{1-\nu_n}{2}\right].
\ee 
In the initial state, the non-zero elements of the matrix are $\Gamma_{2l-1,2l}=-\Gamma_{2l,2l-1}=-{\rm i}$.

The essence of the semiclassical theory of entanglement after a global quench can be formulated in the language of CTQW as follows. The spreading of excitations induced by the sudden quench is described by quantum walks, which are localized on each site at $t=0$ and start to spread out after the quench. 
The entanglement entropy at time $t$ is given by the integrated current $\phi(t)=\int_0^tI(t')dt'$ of quantum walkers from the subsystem to the environment and vice versa. In case of a sudden quench, the Hamiltonian is constant for $t>0$, and the spreading of quantum walks is ballistic. Therefore  $\phi(t)\sim t$, and we have an entanglement entropy that increases linearly in time.

Applying this picture to the model with a fluctuating field, we need to distinguish between two contributions to the current of quantum walkers. First, there is a contribution from the diffusive part of the wave function of CTQW, which is always present. This gives a diffusive current, i.e. $\phi_{d}(t)\sim \sqrt{t}$. 
Second, if there exist stroboscopic eigenmodes, the corresponding ballistic peaks result in a contribution $\phi_b(t)\sim \int^t W(t')dt'\sim t^{1-a}$. Consequently, in the case there are no stroboscopic eigenmodes, we obtain 
\be 
\overline{S(t)}\sim t^{1/2}
\label{S_1/2}
\ee
for long times in an infinite system.
If, however, there exist stroboscopic eigenmodes with a quadratic dispersion, for which $a=1/4$, they give the dominant contribution to the current and lead to
\be 
\overline{S(t)}\sim t^{3/4}.
\label{S_1/4}
\ee
If the stroboscopic eigenmodes have a linear dispersion, thus $a=1/2$, then both $\phi_{d}(t)$ and $\phi_b(t)$
have the same time-dependence, thus the leading behavior is described by Eq.(\ref{S_1/2}). 
Since the shape of the probability distribution is modified due to the presence of stroboscopic eigenmodes one can expect some logarithmic multiplicative correction to the leading behavior.

\begin{figure}
\includegraphics[width=8cm]{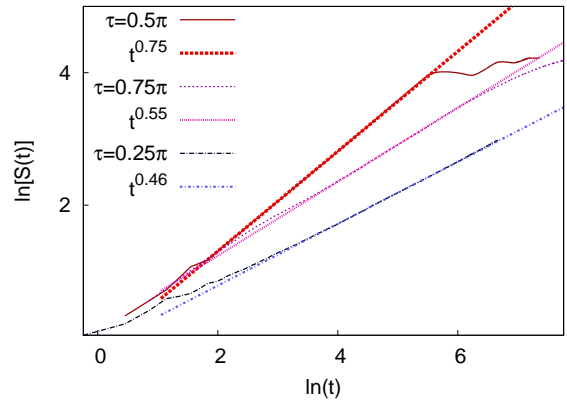}
\caption{\label{ent} (Color online) Time-dependence of the average entanglement entropy calculated numerically for the model with $J_1=J_2=1$, $h_1=-h_2=1$, for $\tau=\pi/4,\pi/2$ and $3\pi/4$. The number of noise realizations was $2000$. The straight lines are linear fits to the data, and the typical deviations of the data from the fitted lines are less than $0.01$ in the fitting range. For  $\tau=\pi/4$ there are no SE-s; for $\tau=\pi/2$ there are ones with a quadratic dispersion; for $\tau=3\pi/4$ there are SE-s with a linear dispersion only.
Note that the deviations from the asymptotic linear form in the large $t$ domain are due to finite-size effects. 
}
\end{figure}

Numerical results for the time-dependence of the entanglement entropy are shown in Fig. \ref{ent}. 
The exponent obtained from finite-time data for the case of no SE-s ($0.47$) is close to the semiclassical
prediction. In the case there are SE-s of linear dispersion only, the numerical value is somewhat larger ($0.55$), while in the case of the presence of SE-s with a quadratic dispersion ($0.75$) the exponent is in agreement with
Eq.(\ref{S_1/4}).  

\subsection{Magnetization}

Finally, we study the relaxation of the magnetization after a quench in the quantum Ising chain with fluctuating transverse fields. After a global quench, it relaxes exponentially for long times\cite{Igloi_11,Rieger_11,Calabrese_11,Calabrese_12,irl}, and this form of time-dependence is correctly reproduced by a semiclassical theory\cite{Igloi_11,Rieger_11}. 
The essence of the latter theory is that quasiparticles emitted from each site after the quench move ballistically and as a kink excitation, or domain wall, flip the spins when they pass by. When several quasiparticles pass a site, then, for odd (even) number of particles, the given spin changes its sign (keeps its value). In the following, we present a simple consideration based on a stochastic process, which is able to predict the functional form of the relaxation. Namely, the orientation of a given spin changes with a rate $I(t)$, which is determined by the current of quasiparticles passing through that site. In the language of CTQW, $I(t)$ is the current of quantum walkers starting out from every site after the quench through the given site.  Then, the probability $p_+(t)$ of the given spin pointing to the positive $x$ direction obeys the master equation
\be 
\frac{d}{dt} 
\begin{bmatrix}
p_+(t) \cr
1-p_+(t)  
\end{bmatrix}
=
\begin{bmatrix}
-I(t)   & I(t) \cr
I(t)    & -I(t) 
\end{bmatrix}
\begin{bmatrix}
p_+(t) \cr
1-p_+(t) 
\end{bmatrix}.
\ee
The solution of this simple equation for the magnetization, $m^x(t)=2p_+(t)-1$, if initially $m^x(0)=1$, is of the form
\be 
m^x(t)=2p_+(t)-1=e^{-2\int_0^tI(t')dt'}.
\label{mxt}
\ee
In the case of a sudden quench, $I(t)$ is constant, and given by the sum of contributions of the quasiparticles as $I=\sum_{k > 0} v_k f_k$, where $v_k=\delta \epsilon_k/\delta k$ is the semiclassical velocity and $f_k$ is the occupation probability of the given mode in the initial state. From this follows, that $m^x(t)=e^{-t/t_r}$, and the relaxation time is given by $t_r=1/(2I)$, which agrees with the semiclassical result in\cite{Igloi_11,Rieger_11}. 

Applying the above considerations for the model with a fluctuating field,
the integrated current, $\phi(t)$, calculated in the previous section has to be inserted in Eq. (\ref{mxt}). This yields a stretched exponential decay of the average magnetization
\be 
\overline{m^x(t)}\sim e^{-ct^{3/4}}
\label{mx34}
\ee
if SE-s with a quadratic dispersion exist, and 
\be 
\overline{m^x(t)}\sim e^{-ct^{1/2}}
\label{mx12}
\ee
otherwise, with a possible logarithmic correction in the case of an SE with linear dispersion.

\begin{figure}
\includegraphics[width=8cm]{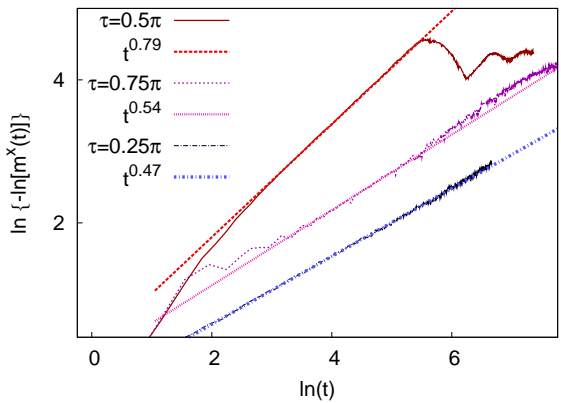}
\caption{\label{magn} (Color online) The same as in Fig. \ref{ent} for the average magnetization. The typical deviations of the data from the fitted lines are less than 0.05 in the fitting range.}
\end{figure}

We have tested the validity of these conjectures by numerically calculating the magnetization $\langle x|\sigma^x(t)|x\rangle$. Using $\langle x|\sigma^x(t)|x\rangle=\langle +|\sigma^x(t)|-\rangle$, where 
$|\pm\rangle=\frac{1}{\sqrt{2}}(|\rightarrow\rightarrow\dots\rightarrow\rangle\pm
|\leftarrow\leftarrow\dots\leftarrow\rangle)$, it can be calculated as Pfaffian, which is given as the square root of a determinant; for the details, see e.g. Ref.\cite{irl}. 
According to the numerical data presented in Fig. \ref{magn}, the average magnetization follows a stretched exponential decay and the estimates of the powers are close to those appear in Eqs. (\ref{mx34}) and (\ref{mx12}) with small deviations seen also for other quantities.  


\section{Discussion}
\label{discussion}
We have studied in this work the nonequilibrium relaxation dynamics of the Ising chain in a fluctuating transverse field. As the model can be reformulated as a Majorana chain with quadratic terms, its true dimensionality is $2L$ rather than $2^L$, and, using this, we have pointed out a close relationship between the dynamics of the model and continuous-time quantum walks. Namely, the equations of motion of Majorana operators, from which all observables can be built up, in the Heisenberg picture are formally identical to a Schr\"odinger equation of a particle on a chain of length $2L$, i.e. a one-dimensional continuous-time quantum walk. 
The linear time-dependence of the entanglement entropy and logarithmic magnetization in the case of a sudden quench are then intimitely related to the well-known ballistic spreading of the CTQW associated with the model. 

In the case of a fluctuating transverse field, the associated one-particle model will be a noisy (or random) CTQW. The external noise, in general, is known to destroy quantum interference, and CTRW crosses over to a classical random walk with the well-known diffusive dynamics\cite{decoherence,kendon}.  
As we have shown by a semiclassical reasoning and confirmed by numerical calculations, the diffusive spreading leads to a square-root time-dependence of the entanglement entropy and logarithmic magnetization in the fluctuating model. 
Although we have carried out calculations in a concrete integrable model, 
we conjecture the square-root increase of the entanglement entropy to be generally valid for one-dimensional quantum systems subject to an external noise, in the pure version of which entropy growth is realized by the propagation of entangled quasiparticle excitations. 

For the particular case of a dichotomous noise composed of segments of constant duration, we have found that coherence is not completely destroyed. Namely, so called stroboscopic eigenmodes can exist, which are common eigenmodes of both clean evolution operators, and these appear as ballistically propagating but algebraically decaying peaks in the wave function of the associated CTQW. If SE-s with a quadratic dispersion exist, the contribution of which is dominant over the diffusive one, the power $1/2$ in the above laws changes to $3/4$. 
We stress, however, that this phenomenon arises only for the particular form of the noise with segments of constant duration, and not for a general dichotomous Markov noise or other noise with a random duration of segments. 
Even a small noise in the duration $\tau$ or in the transverse fields is expected to introduce a finite lifetime for stroboscopic eigenmodes, which could be obtained by extending the calculations of the Appendix, but this is out of the scope of the present work. 

Further directions in connection with stochastic noise that could be explored are, among others, local quench dynamics instead of a global one considered in this work, or inclusion of spatial inhomogeneity, which, in the absence of noise, gives rise to an ultra-slow dynamics in the critical model and localization otherwise. These are left for future research.

\begin{acknowledgments}

We thank discussions with J\'anos Asb\'oth and Zolt\'an Zimbor\'as. 
This work was supported by the Hungarian Scientific Research Fund under Grant No. K109577, in part by the National Science Foundation under Grant No. NSF PHY11-25915, 
and partially funded by the T\'AMOP-4.2.2.B-15/1/KONV-2015-0006
project, which is supported by the European Union and co-financed by
the European Social Fund.

\end{acknowledgments}

\appendix

\section{Decay rate of modes}

\label{app:decay}

To calculate the probability $P_k^+(t)$ defined in Eq. (\ref{Pk+}), notice that the state $|f(t)\rangle$ remains in the two-dimensional subspace spanned by $|f^+_{k,A}\rangle$ and $|f^-_{k,A}\rangle$.
The unitary evolution operators of system $A$ and $B$ are represented in this basis by the matrices 
\be 
U_A= 
\begin{bmatrix}
e^{-{\rm i}\epsilon_{k,A}^+\tau}   & 0 \cr
0    &  e^{{\rm i}\epsilon_{k,A}^+\tau}
\end{bmatrix}, 
\quad U_B= 
\begin{bmatrix}
\omega_k^*   & -\gamma_k \cr
\gamma_k    &  \omega_k
\end{bmatrix},
\label{Umatrices}
\ee
where  $\omega_k=\cos(\epsilon_{k,B}^+\tau)+{\rm i}\sin(\epsilon_{k,B}^+\tau)\cos(2\Delta_k)$ and $\gamma_k=\sin(\epsilon_{k,B}^+\tau)\sin(2\Delta_k)$. 

Denoting the vector representing the state at time $t_n$ by $[F_1(t_n),F_2(t_n)]^T$, and introducing $F_{ij}^{(n)}\equiv \overline{F_i^*(t_n)F_j(t_n)}$, $i,j=1,2$, we can write, similarly to Eq. (\ref{directav}), 
\be
F_{ij}^{(n)}=\sum_{k,l=1}^{2}\frac{1}{2}\{[U_A^*]_{ik}[U_A]_{jl}+[U_B^*]_{ik}[U_B]_{jl}\}F_{kl}^{(n-1)}.
\label{Fav}
\ee
Defining a four-component vector as $F^{(n)}=[F_{11}^{(n)},F_{12}^{(n)},F_{21}^{(n)},F_{22}^{(n)}]^T$, the r.h.s. of Eq. (\ref{Fav}) amounts to a multiplication of $F^{(n-1)}$ by the $4\times4$ matrix 
$\overline{U}\equiv\frac{1}{2}(U_A^*\otimes U_A+U_B^*\otimes U_B)$: 
\be 
F^{(n)}=\overline{U}F^{(n-1)}.
\ee
The long-time behavior of $P_k^+(t_n)=F_{11}^{(n)}$ is determined by the spectrum of $\overline{U}$. It has a unit eigenvalue for all $k$, which corresponds to the stationary state $\lim_{n\to\infty}F^{(n)}=[\frac{1}{2},0,0,\frac{1}{2}]^T$, while  
the asymptotic decay of $P_k^+(t_n)$ is determined by the eigenvalue with the second largest modulus $r_k$, as 
\be 
P_k^+(t_n)-P_k^+(\infty)\sim r_k^n\sim e^{-t_n/\tau_k}, 
\ee
where the lifetime $\tau_k$ of the mode is $\tau_k=\tau/|\ln r_k|$.

\end{document}